\documentclass[aps,prm,superscriptaddress,twocolumn]{revtex4-1}
\usepackage{graphics}
\usepackage{graphicx} % Include figure files
\usepackage{dcolumn}% Align table columns on decimal point
\usepackage{bm}% bold math
\usepackage{epstopdf}
\usepackage[usenames, dvipsnames]{color}
\usepackage{amsmath}
\usepackage{times}

\begin{document}

% Use the \preprint command to place your local institutional report
% number in the upper righthand corner of the title page in preprint mode.
% Multiple \preprint commands are allowed.
% Use the 'preprintnumbers' class option to override journal defaults
% to display numbers if necessary
%\preprint{}

%Title of paper
\title{Relationship between A-site Cation and Magnetic Structure in 3$d$-5$d$-4$f$ Double Perovskite Iridates $Ln$$_2$NiIrO$_6$ ($Ln$=La, Pr, Nd)}

% repeat the \author .. \affiliation  etc. as needed
% \email, \thanks, \homepage, \altaffiliation all apply to the current
% author. Explanatory text should go in the []'s, actual e-mail
% address or url should go in the {}'s for \email and \homepage.
% Please use the appropriate macro foreach each type of information

% \affiliation command applies to all authors since the last
% \affiliation command. The \affiliation command should follow the
% other information
% \affiliation can be followed by \email, \homepage, \thanks as well.
%\author{}
%\email[]{Your e-mail address}
%\homepage[]{Your web page}
%\thanks{}
%\altaffiliation{}
%\affiliation{}

\author{T.~Ferreira}
\affiliation{Materials Science and Technology Division, Oak Ridge National Laboratory, Oak Ridge, TN 37831.}
\affiliation{Department of Chemistry and Biochemistry, University of South Carolina, Columbia, SC 2920.}

\author{S.~Calder}
\email{caldersa@ornl.gov}
\affiliation{Neutron Scattering Division, Oak Ridge National Laboratory, Oak Ridge, Tennessee 37831, USA.}

\author{D.~S.~Parker}
\affiliation{Materials Science and Technology Division, Oak Ridge National Laboratory, Oak Ridge, TN 37831.}

\author{M.~H.~Upton}
\affiliation{Advanced Photon Source, Argonne National Laboratory, Argonne, Illinois 60439, USA.}

\author{A.~S.~Sefat}
\affiliation{Materials Science and Technology Division, Oak Ridge National Laboratory, Oak Ridge, TN 37831.}

\author{H.-C.~zur Loye}
\email{zurloye@mailbox.sc.edu}
\affiliation{Department of Chemistry and Biochemistry, University of South Carolina, Columbia, SC 2920.}
%Collaboration name if desired (requires use of superscriptaddress
%option in \documentclass). \noaffiliation is required (may also be
%used with the \author command).
%\collaboration can be followed by \email, \homepage, \thanks as well.
%\collaboration{}
%\noaffiliation

\begin{abstract}	
We report a comprehensive investigation of $Ln_2$NiIrO$_6$ ($Ln$ = La, Pr, Nd) using thermodynamic and transport properties, neutron powder diffraction, resonant inelastic x-ray scattering, and density functional theory (DFT) calculations to investigate the role of A-site cations on the magnetic interactions in this family of hybrid 3$d$-5$d$-4$f$ compositions. Magnetic structure determination using neutron diffraction reveals antiferromagnetism for La$_2$NiIrO$_6$, a collinear ferrimagnetic Ni/Ir state that is driven to long range antiferromagnetism upon the onset of Nd ordering in Nd$_2$NiIrO$_6$, and a non-collinear ferrimagnetic Ni/Ir sublattice interpenetrated by a  ferromagnetic Pr lattice for Pr$_2$NiIrO$_6$. For Pr$_2$NiIrO$_6$ heat capacity results reveal the presence of two independent magnetic sublattices and transport resistivity indicates insulating behavior and a conduction pathway that is thermally mediated. First principles DFT calculation elucidates the existence of the two independent magnetic sublattices within Pr$_2$NiIrO$_6$ and offers insight into the behavior in La$_2$NiIrO$_6$ and Nd$_2$NiIrO$_6$. Resonant inelastic x-ray scattering is consistent with spin-orbit coupling splitting the t$_{2g}$ manifold of octahedral Ir$^{4+}$ into a J$\rm _{eff}$ = $\frac{1}{2}$ and J$\rm _{eff}$ = $\frac{3}{2}$ state for all members of the series considered. 
\end{abstract}

%\maketitle must follow title, authors, abstract, \pacs, and \keywords
\maketitle

\section{Introduction}

Perovskites are one of the most studied solid-state materials due to their modular structure allowing for the incorporation of a wide range of elements, within the limitations outlined by the Goldschmidt tolerance factor  \cite{Goldschmidt, Lufaso:br0106,Lufaso:ws5032,doi:10.1021/cm00040a007}. The ability to stabilize a wide variety of elements with different, and often competing, physical properties within the same material makes the perovskite structure a model system to study a rich diversity of magnetic and electronic properties  \cite{doi:10.1021/acs.inorgchem.7b01086, DAVIS2004413, doi:10.1021/ja411713q, doi:10.1021/acs.cgd.6b00121,doi:10.1021/ic048637x, KAYSER2017114,doi:10.1021/cm0716708, PhysRevB.79.224428,doi:10.1021/acs.chemmater.6b00755,doi:10.1021/ic051045+,doi:10.1021/ja407342w,PhysRevB.92.094435, doi:10.1021/acs.chemmater.6b00254, MUGAVEROIII2010465,doi:10.1021/acs.inorgchem.8b02557,PhysRevB.97.184407,PhysRevB.94.235158,doi:10.1021/ic503086a, doi.org/10.1002/ejic.201500569,doi:10.1021/acs.chemmater.6b04983}. Hybrid 3$d$-5$d$(4$d$) based materials that adopt the perovskite structure type host an array of physical properties originating from a delicate balance of interactions. For example, unpaired 3$d$ electrons strongly correlate to 2$p$ oxygen electrons in a perovskite lattice, often resulting in technologically useful properties such as ferromagnetism  \cite{ISI:000073241200046}, ferroelectricity  \cite{ISI:A1992JC58300046}, and multiferroicism  \cite{doi:10.1021/acs.inorgchem.7b01086}. By contrast, the greater orbital extent of heavier 5$d$ elements, weaker electron correlation strength, and stronger spin-orbit coupling (SOC) can lead to metal-insulator transitions \cite{NaOsO3Calder}, topological insulators  \cite{NaturePesin}, superconductivity  \cite{ISI:000367835400016} and a split of the t$_{2g}$ manifold into a J$\rm _{eff}$ = $\frac{1}{2}$ and J$\rm _{eff}$ = $\frac{3}{2}$ state, as observed in Sr$_2$IrO$_4$  \cite{KimScience}, that can lead to new routes to Mott and other exotic insulating states  \cite{calder2015_cd227,PhysRevB.87.155136,PhysRevB.92.235109,PhysRevB.91.214433,NatPhysChun,PhysRevB.92.121113,PhysRevLett.110.027002,PhysRevLett.108.127203,PhysRevLett.109.027204,KimScience,PhysRevLett.102.017205,zaac.200900039}. Hybrid perovskites containing both 3$d$ and 5$d$ elements have been reported to exhibit a wide range of properties characteristic of both 3$d$ and 5$d$ containing oxides, in addition to extremely high magnetic ordering temperatures, (Curie temperature of T$\rm _c$ = 725 K) such as that observed in Sr$_2$CrOsO$_6$, further motivating the study of perovskites as a host lattice to investigate the balance of competing interactions.

Compared to the single perovskite (ABO$_3$) system with only one B site, the double perovskite (A$_2$BB'’O$_6$) allows for two crystallographically unique sites on which up to three magnetic ions may reside. Most studies of double perovskites limit the number of magnetic cations to one or two, often on the B and B' site for ease of study, although exceptions do exist  \cite{doi:10.1021/acs.cgd.6b00121,doi:10.1021/acs.chemmater.6b00755}. This allows for the possibility of studying the interaction between superexchange (B-O-B') and super-superexchange interactions (B-O-B'-O-B), such as that studied in Ca$_2$MOsO$_6$ (M = Co, Ni)   \cite{doi:10.1021/acs.chemmater.6b00254,doi:10.1021/ic051045+}. There it was demonstrated that strong antiferromagnetic coupling between Os and Co/Ni stabilize the ferrimagnetic ground state, indicating strong superexchange interactions, and weak super-superexchange interactions. Interestingly, the chemical substitution of nonmagnetic Ca in these materials for Sr results in Sr$_2$CoOsO$_6$, which has been shown to exhibit strong super-superexchange interactions (Os-O-Co-O-Os and Co-O-Os-O-Os) resulting in two interpenetrating antiferromagnetic magnetic sublattices  \cite{doi:10.1021/ja407342w}. These sublattices have independent magnetic ordering temperatures (Os: T$\rm _N$ = 108 K; Co: T$\rm _N$ = 70 K) and distinct magnetic propagation vectors (Os: k=($\frac{1}{2}$,$\frac{1}{2}$,0); Co: k=($\frac{1}{2}$,0,$\frac{1}{2}$) )  \cite{doi:10.1021/ja407342w}, in direct contrast to the nearly isostructural and isovalent Ca$_2$CoOsO$_6$ analog. The subtle structural change associated with substitution of Ca for Sr resulted in a drastic change in superexchange strength, magnetic ordering temperature, and the nature of the long range magnetic order (ferrimagnetic Ca$_2$CoOsO$_6$ and antiferromagnetic Sr$_2$CoOsO$_6$), exemplifying how sensitive these hybrid perovskites are to chemical changes  \cite{MORROW201646}.

The Kanamori-Goodenough rules  \cite{KANAMORI195987} have provided a set of semi-empirical guidelines to understand the complex relationship between superexchange interactions and magnetic order in condensed matter systems. These rules provide a method for determining the sign of superexchange interactions, predicting antiferromagnetic order for linear M-X-M interactions (where X is a bridging anionic unit such as a chalcogenide or halide) and ferromagnetic order for 90$^{\circ}$ M-X-M interactions. Although these rules have been shown to successfully predict superexchange interactions for perovskites, poor energetic overlap between magnetic cations, such as those in mixed 3$d$-5$d$ oxides, can lead to violations of these rules. One such example is the hybrid 3$d$-5$d$ double perovskite Sr$_2$FeOsO$_6$, \cite{PhysRevLett.111.167205,PhysRevB.98.214422} in which the bent Os-O-Fe superexchange interaction in the $ab$-plane exhibited antiferromagnetic order, and these bonds exhibited ferromagnetism in the $c$-axis despite the 180$^{\circ}$ Os-O-Fe bond angle. Exceptions such as these continue to motivate the detailed study of hybrid 3$d$-5$d$ complex oxides, and serve as a motivating factor for this work, which extends to the rarely studied 3$d$-5$d$-4$f$ compositions. 

Here we report a comprehensive investigation of $Ln$$_2$NiIrO$_6$ ($Ln$ = La, Pr, Nd). We begin with measurements of all compounds with neutron powder diffraction and resonant inelastic x-ray scattering (RIXS) to determine the magnetic structure and explore how SOC affects the t$_{2g}$ manifold of the Ir ion. The remainder of the manuscript focuses on  Pr$_2$NiIrO$_6$  using thermodynamic, transport property and density functional theory (DFT) calculations. The results allow insights into the role superexchange plays in these scarcely studied hybrid 3$d$-5$d$-4$f$ compositions with variable A site cations. These materials were previously reported by some of the authors of this manuscript  \cite{doi:10.1021/acs.cgd.6b00121} and this study seeks to elucidate the magnetic structure of all three compositions. Several magnetic ordered phases are observed as the different magnetic ions order. The presence of independent magnetic sublattices in Pr$_2$NiIrO$_6$ is explored in detail. This approach allows us to go beyond the Kanamori-Goodenough rules to determine the varied magnetic interactions and ground states in these related materials as the rare earth ion is altered and the temperature is tuned.

\section{Experimental Details}

\subsection{Sample synthesis}

$Ln_2$O$_3$ (Alfa Aesar 99.99$\%$) and Pr$_6$O$_{11}$ (Alfa Aesar 99.9$\%$) were all heated in air at 1000$^{\circ}$C in a tube furnace overnight to remove any possible hydroxide or carbonate impurities. Pr$_6$O$_{11}$ (Alfa Aesar, 99.99$\%$) was reduced to Pr$_2$O$_3$ under 5$\%$ hydrogen at 1000$^{\circ}$C in a tube furnace overnight. NiO (Sigma Aldrich, 99.999$\%$) and Ir powder (Engelhard, 99.9995$\%$) were used as received. Polycrystalline samples of $Ln$$_2$NiIrO$_6$ were prepared by intimately grinding $Ln$$_2$O$_3$, Ni, and Ir metal in stoichiometric amounts and heating the resultant powder in air in an alumina crucible with a loose fitting lid. The samples were heated to 800$^{\circ}$C for 72 hours, 900$^{\circ}$C for 72 hours, and then 975$^{\circ}$C for 168 hours with intermediate grindings in a programmable furnace. For Pr$_2$NiIrO$_6$, an additional heating at 1025$^{\circ}$C for 96 hours with intermediate grindings was necessary.

\subsection{Physical property measurements} 

Temperature dependent heat capacity was measured using a Quantum Design physical property measurement system (PPMS) on polycrystalline powder of Pr$_2$NiIrO$_6$ that were pressed into a pellet and sintered at 400$^{\circ}$C for 72 hours.
The electrical resistance of pressed and sintered pellets cut into a rectangular shape was recorded as a function of temperature by the four-probe method. Silver paint electrodes using platinum wires were used as contact points. The temperature was controlled from 380 K down to 1.8 K using a Quantum Design PPMS.

\subsection{Neutron powder diffraction}

Neutron diffraction measurements were performed on 5 gram samples of $Ln$$_2$NiIrO$_6$ at Oak Ridge National Laboratory on the HB-2A Powder diffraction instrument at the High Flux Isotope Reactor (HFIR) \cite{Garlea2010, doi:10.1063/1.5033906}. Measurements were performed with the samples loaded into 1 mm Al annular cans to reduce neutron absorption from the Ir ion. The outer diameter of the sample cans were 15 mm. A wavelength of 2.41 $\rm \AA$  was selected with a vertically focusing germanium monochromator on the Ge(113) reflection. Data were collected over a 2$\theta$ angular range of 5$^{\circ}$ - 130$^{\circ}$ in steps of 0.05$^{\circ}$. The detector efficiency was normalized with a vanadium measurement. The La$_2$NiIrO$_6$ and Nd$_2$NiIrO$_6$ samples were cooled in a top-loading closed cycle refrigerator (CCR) to reach 4 K and a $^4$He cryostat was used for Pr$_2$NiIrO$_6$ to get to the lower temperature of 1.5 K. FullProf  was utilized for the Rietveld refinement and determination of the propagation vectors (k vectors) \cite{Fullprof}. The magnetic space groups were determined using the Bilbao Crystallographic Server \cite{Bilbao_Mag, ISI:000292661600002}. Representational analysis was also used during the magnetic structure determination process with SARAh \cite{sarahwills}. See Supplemental Material at \cite{SuppInfo} for the mcif files of the determined magnetic structures.

\subsection{Resonant Inelastic X-ray Scattering}

RIXS was carried out on the MERIX spectrometer, sector-27 at the Advanced Photon Source (APS) \cite{SHVYDKO2013140}. The incident energy was tuned to the Ir L$_3$-edge (11.215 keV) resonant edge to enhance the Ir scattering. The inelastic energy was measured with the use of a Si(844) analyzer. The energy resolution was determined to be 35 meV at full width half maximum (FWHM), based on fitting the quasi-elastic line to a charge peak. The scattering plane and incident photon polarization were both horizontal, i.e.~$\pi$ incident polarization, with the incident beam focused to a size of 40 $\times$ 25 $\mu$m$^2$ (H$\times$V) at the sample position. To minimize elastic scattering measurements were performed with 2$\theta$ at 90$^{\circ}$ in horizontal geometry. All measurements were performed on powder samples mounted onto an Al block sealed with Kapton paper with space for all three samples in a custom mount. The temperature was controlled with a CCR and measurements taken at 5 K, 30 K and 150 K to cover the different regions of magnetic ordering in the materials.

\subsection{First Principles Calculations} First principles calculations were performed using the all-electron linearized augmented planewave (LAPW) DFT code WIEN2K  \cite{wien2k}, within the generalized gradient approximation of Perdew, Burke and Ernzerhof \cite{pbe}. LAPW sphere radii of 1.62, 1.98, 1.98 and 2.35 Bohr were used respectively for Oxygen, Nickel, Iridium and Praseodymium, respectively, with an RK$\rm _{max}$ value of 8.0 employed. Here RK$\rm _{max}$ is the product of the smallest sphere radius (in this case Oxygen) and the largest plane-wave expansion wavevector. All calculations used an optimized structure, with the lattice constants and space group taken from the experimental measurement and all internal coordinates not dictated by symmetry relaxed within a ferromagnetic Pr-Ni configuration (note that in this case Ir carries a small negative moment). Sufficient numbers of k-points (generally between 200 and 600 in the full Brillouin zone) to describe the magnetic order were used for all calculations. For the detailed magnetic calculations (not the optimization), a U value of 5 eV was applied to the Pr 4f orbitals. This value corresponds with that chosen in recent work on Pr-containing transition metal, perovskite oxides \cite{rezaiguia2017gga+}. We also include straight GGA results as this provides insight regarding the effect of the Hubbard U on the exchange energetics.

\section{Results and Discussion}

\subsection{Magnetic Structure Determination}

Neutron powder diffraction measurements were performed on all $Ln$$_2$NiIrO$_6$ materials to determine the magnetic structure in different temperature regimes. The crystal structures were previously determined with single crystal x-ray diffraction to be $P2_1/n$ ($\# 14$) \cite{doi:10.1021/acs.cgd.6b00121}.

\subsubsection{La$_2$NiIrO$_6$}

\begin{table}[htb]
	\begin{tabular}{|c|c|c|c|c|}
		\hline
		& x & y & z & Site \\
		\hline
		La & 0.008(1) &  0.546(5) &  0.753(2) & 4e \\
		
		Ir & 0 & 0 & 0 & 2a \\
		
		Ni & 0 & 0 & 0.5 & 2b \\
		%	\hline
		O1 &  0.084(1)  & 0.019(8) & 0.260(2) & 4e \\
		%	\hline
		O2 & 0.211(4) &  0.280(2) & -0.047(1) & 4e \\
		%	\hline
		O3 &  0.207(4) &  0.305(2) & 0.540(2) & 4e \\
		\hline
	\end{tabular}
	\caption{Crystal structure of La$_2$NiIrO$_6$ at 100 K from neutron refinement in the $P2_1/n$ space group with $a$=5.566(2)$\rm \AA$, $b$=5.630(2)$\rm \AA$, $c$=7.888(3)$\rm \AA$, $\beta$=90.09(2)$^{\circ}$.}
	\label{La_table}
\end{table}

%trim is left, bottom, right, top
\begin{figure}[tb]
	\centering         
	\includegraphics[trim=0cm 13.4cm 0cm 0cm,clip=true, width=1.0\columnwidth]{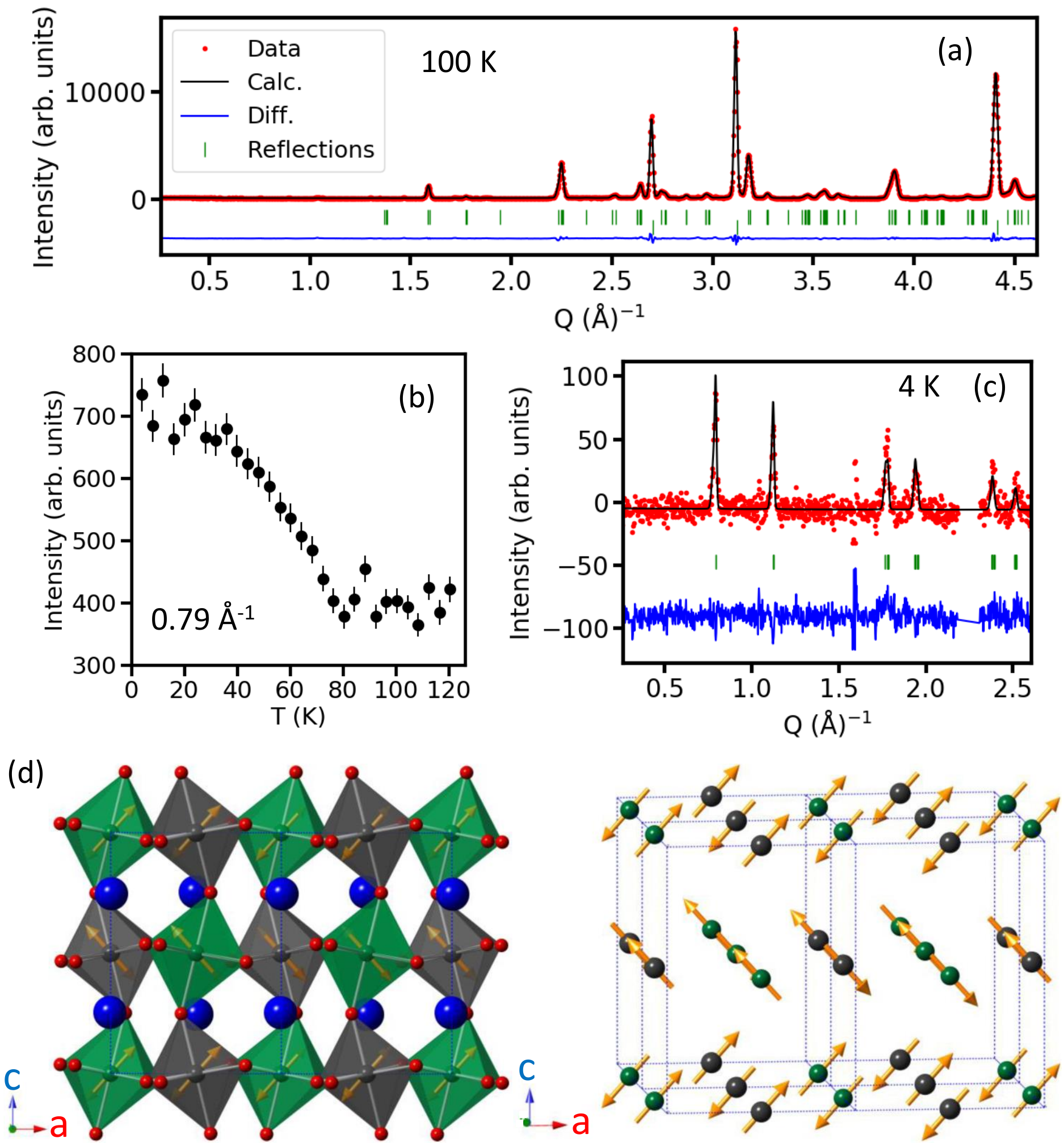}           
	\caption{\label{NPD_La} Magnetic structure of La$_2$NiIrO$_6$. (a) Refinement of neutron powder diffraction data at 100 K to the $P2_1/n$ crystal structure (upper tick marks). Lower Tick marks correspond to the Al scattering from the sample holder. (b) Intensity of the reflection at 0.79 $\rm \AA^{-1}$ as a function of temperature. (c) Magnetic structure model fit to the intensity obtained by subtracting the 100 K neutron diffraction data from the 4 K measurement. (d) Polyhedral representation of the magnetic and nuclear structure of La$_2$NiIrO$_6$ and magnetic-atom only representation  with La (blue), Ni (green), Ir (grey) and O (red) atoms shown. The non-magnetic unit cell is outlined with the blue dashed line. The magnetic unit cell is doubled along the $a$ and $b$-axis.}
\end{figure} 

 We first begin with the magnetic structure determination of  La$_2$NiIrO$_6$, that is expected to only contain 3$d$ (Ni$^{2+}$) and 5$d$ (Ir$^{4+}$) magnetic ion ordering and a non-magnetic $4f$ ion (La$^{3+}$). La$_2$NiIrO$_6$ was reported to undergo an antiferromagnetic transition around 75 K, with indications of further magnetic anomalies within this phase  \cite{doi:10.1021/acs.cgd.6b00121}. A powder sample of La$_2$NiIrO$_6$ was measured at four different temperatures: 4 K, 40 K, 65 K, and 100 K. This allowed the anomalies in the reported SQUID measurements  \cite{doi:10.1021/acs.cgd.6b00121} to be explored and disentangle the evolution of any magnetic ordering. The 100 K measurement is above the highest observed magnetic transition and was used to obtain a structural model in the paramagnetic phase based on the previously reported structure of $P2_1/n$, shown in Fig.~\ref{NPD_La}(a) and Table \ref{La_table}. No impurities were detected in the neutron data. Upon cooling below 80 K, the temperature regime in which magnetic order is expected for La$_2$NiIrO$_6$, the presence of additional intensity was observed. The intensity change at a forbidden nuclear position was followed in Fig.~\ref{NPD_La}(b) to track the onset of magnetic ordering. This indicated a magnetic ordering below T$\rm _N$=80 K, consistent with the SQUID data  \cite{doi:10.1021/acs.cgd.6b00121}. The same magnetic reflections were present at 65 K, 40 K, and 4 K with no change indicative of further magnetic transitions within the resolution of the present measurements. Further measurements on single crystals or with higher resolution will be of interest to probe these subtle changes observed in Ref.\cite{doi:10.1021/acs.cgd.6b00121}. 
 
 A propagation vector of k = ($\frac{1}{2}$, $\frac{1}{2}$, 0) was determined from the positions of the magnetic reflections. Using the Bilbao Crystallographic Server and non-magnetic space group $P2_1/n$ (non-standard setting) with the determined k-vector gives the $P_S$-1 ($\#2.7$) magnetic space group as the only maximally allowed structure with non-zero moments. The direction of the moments is unconstrained in this model. Given the number of variables, powder averaging inherent in the data and small contribution from the Ir ion we attempted to limit the spin directions to uncover the dominant component. Confining the spins to the $a$-axis produced the most reasonable agreement of any of the trial $a$,$b$,$c$ directions to the data with an R$\rm _{mag}$ value of 22.1. Allowing the spins to have a component along the $c$-axis further increased the agreement to the data with an R$\rm _{mag}$ value of 6.25. This model is shown in Fig.~\ref{NPD_La}(d). When the moments were allowed to freely refine along all directions the $b$-axis produced a value with a large error within zero, distinct from the $a$ and $c$ axis. As such we present a magnetic model for La$_2$NiIrO$_6$ with only $a$-$c$ spin components, however we cannot rule out a $b$-axis component. The refined moment values in our model were 1.53(5)$\rm \mu_B/Ni^{2+}$ with components  ($\rm m_a$,$\rm m_b$,$\rm m_c$)= (1.0,0,1.1) and 0.17(3)$\rm \mu_B/Ir^{4+}$ with components  ($\rm m_a$,$\rm m_b$,$\rm m_c$)= (0.12,0,0.13). We note that the low moment of Ir$^{4+}$ is beyond the typical limit for this measurement and therefore is presented as the best fit model. The errors from the Rietveld refinement are likely an underestimation and we cannot rule out the Ir$^{4+}$ having a zero moment or the Ir and Ni sublattices ordering at different temperatures. Further measurements sensitive to the Ir ion, such as resonant x-ray scattering, would be of interest, as would measurements on crystals to determine the spin direction of all the moments. 
 
 \subsubsection{Nd$_2$NiIrO$_6$}

\begin{table}[htb]
	\begin{tabular}{|c|c|c|c|c|}
		\hline
		& x & y & z & Site \\
		\hline
		Nd & 0.015(2) &  0.565(1) &  0.565(1) & 4e \\
		
		Ir & 0 & 0 & 0 & 2a \\
		
		Ni & 0 & 0 & 0.5 & 2b \\
		%	\hline
		O1 &  0.099(2) &  0.029(2) & 0.267(3) & 4e \\
		%	\hline
		O2 & 0.184(4) & 0.283(5) & -0.057(3) & 4e \\
		%	\hline
		O3 &  0.202(4) & 0.312(5) &  0.548(4) & 4e \\
		\hline
	\end{tabular}
	\caption{Crystal structure of Nd$_2$NiIrO$_6$ at 150 K from neutron refinement in the $P2_1/n$ space group with $a$=5.429(7)$\rm \AA$, $b$=5.682(7)$\rm \AA$, $c$=7.753(9)$\rm \AA$, $\beta$=90.17(2)$^{\circ}$.}
	\label{Nd_table}
\end{table}

%trim is left, bottom, right, top
\begin{figure}[htb]
	\centering         
	\includegraphics[trim=0cm 1.4cm 0cm 0cm,clip=true, width=1.0\columnwidth]{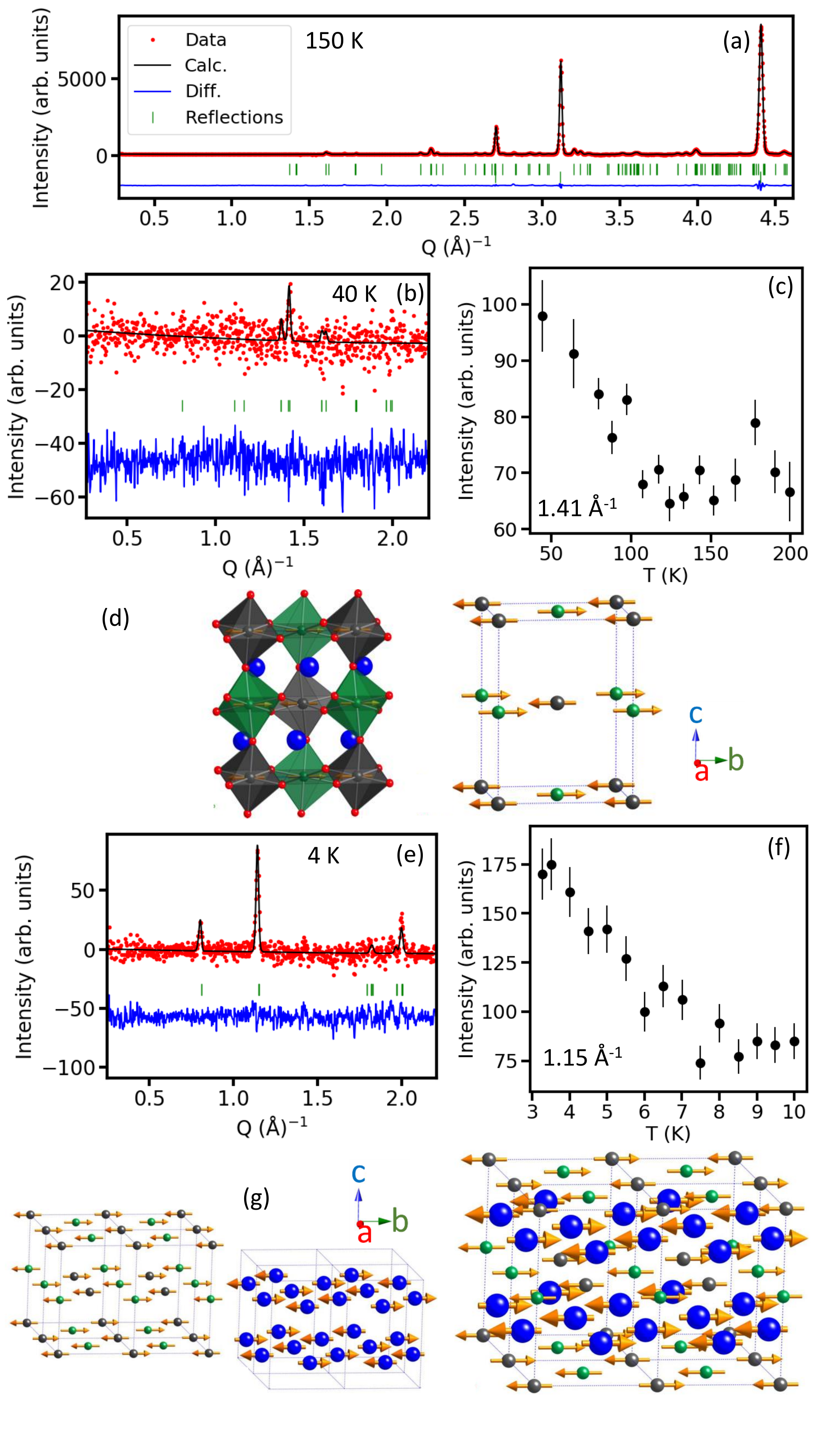}           
	\caption{\label{NPD_Ni}  Magnetic structure of Nd$_2$NiIrO$_6$. (a) Refinement of the 150 K neutron diffraction pattern  to the crystal structure (upper reflections). Lower reflections correspond to Al sample holder scattering. (b) Magnetic structure model obtained by subtracting the 150 K neutron pattern from the 40 K data. (c) Intensity of the reflection at 1.41 $\rm \AA^{-1}$ as a function of temperature. (d) 40 K magnetic structure model with Nd (blue), Ni (green) and Ir (grey) atoms shown. (e) Magnetic structure model obtained by subtracting the 150 K neutron pattern from the 4 K data. (f) Intensity of the reflection at 1.15 $\rm \AA^{-1}$ as a function of temperature. (g) 4 K magnetic structure for Ni/Ir ions (left), Nd ions (middle) and all magnetic ions (right). One magnetic unit cell is shown with the dashed lines correspond to the non-magnetic unit cell.}
\end{figure}

We now turn to the compositions with magnetic 3$d$-5$d$-4$f$ ions. Nd$_2$NiIrO$_6$ was reported to have a ferromagnetic-like transition around 125 K with a further anomaly at 6 K consistent with antiferromagnetic interactions, based on SQUID measurements  \cite{doi:10.1021/acs.cgd.6b00121}. To follow the magnetic structure we therefore collected neutron diffraction measurements  at 4, 40, and 150 K. The high temperature measurement, shown in Fig.~\ref{NPD_Ni}(a) and Table \ref{Nd_table}, was used to confirm purity and obtain a non-magnetic structural model in the paramagnetic regime. This was refined with the $P2_1/n$ space group. The 40 K measurement revealed additional scattering, which is shown in Fig.~\ref{NPD_Ni}(b) where the 150 K data has been subtracted from the 40 K data. The intensity of the scattering at 1.41 $\rm \AA^{-1}$ was followed as a function of temperature, see Fig.~\ref{NPD_Ni}(c). The increase in intensity is consistent with the predicted magnetic ordering at 125 K from bulk data  \cite{doi:10.1021/acs.cgd.6b00121}. The additional scattering could be indexed to a propagation vector of k = (0, 0, 0). Given this k vector and $P2_1/n$ symmetry of the nuclear structure gives four maximally allowed magnetic space groups: $P2_1'/c' (\#14.79)$, $P2_1/c' (\#14.78)$, $P2_1'/c (\#14.77)$ and $P2_1/c (\#14.75)$. Magnetic space groups $\#14.77$  and $\#14.78$ only allow moments on the Nd ion and could be discarded. The remaining two magnetic space groups do not constrain the moments to any fixed axis. The best fit to the data was obtained with a ferrimagnetic arrangement of Ni and Ir in the $b$ axis in the magnetic space group $P2_1/c (\#14.75)$. Magnetic moments of 1.71(2)$\rm \mu_B$ (Ni) and 0.32(7)$\rm \mu_B$ (Ir) were determined, corresponding to R$\rm _{mag}$ of 25.7. The higher agreement index of this fit compared to previous refinements is due to the weaker intensity and reduced number of the magnetic reflections in this phase. Attempts to improve this R$\rm _{mag}$ by introducing components away from the $b$-axis did not appreciably improve the fit. A ferromagnetic model is additionally in agreement with the data, however the ferrimagnetic model presented is more consistent with bulk measurements previously reported  \cite{doi:10.1021/acs.cgd.6b00121}. We note again that the low moment of Ir$^{4+}$ is beyond the typical limit for this measurement and therefore is presented as the best fit model.

Upon cooling to 4 K, additional magnetic reflections appeared in the diffraction pattern, shown in Fig.~\ref{NPD_Ni}(e) for the difference between the 4 K data and the 150 K data. The intensity at the most intense reflection position was followed as a function of temperature in Fig.~\ref{NPD_Ni}(f). This revealed the onset of magnetic ordering below 7 K, consistent with reported SQUID results  \cite{doi:10.1021/acs.cgd.6b00121}. The magnetic reflections observed at 4 K were indexed to a k=($\frac{1}{2}$, $\frac{1}{2}$, 0) propagation vector within the non-magnetic space group $P2_1/n$. Only one maximal  magnetic space group allows moments for Ni/Ir, as well as Nd: $P_S$-1. A magnetic model with the spins still confined to the $b$-axis but in an antiferromagnetic arrangement for all the ions yields the best fit to the data. Magnetic moments of 2.20(4)$\rm \mu_B/Nd^{3+}$ , 1.27(4)$\rm \mu_B/Ni^{2+}$ and 0.32(5)$\rm \mu_B/Ir^{4+}$  were determined.

The magnetic behavior of Nd$_2$NiIrO$_6$ upon cooling is therefore characterized as first undergoing ferrimagnetic ordering of the Ni/Ir ions with the magnetic order keeping the unit cell size unaltered. Then only at the low temperature of 7 K does the Nd ion order along with a change in the ordering of the Ni/Ir magnetic order to antiferromagnetic to create a magnetic unit cell doubled in size along the $a$ and $b$ axis.

\subsubsection{Pr$_2$NiIrO$_6$}

\begin{table}[htb]
	\begin{tabular}{|c|c|c|c|c|}
		\hline
		& x & y & z & Site \\
		\hline
		Pr & 0.012(3) & 0.560(1) & 0.754(5) & 4e \\
		
		Ir & 0 & 0 & 0 & 2a \\
		
		Ni & 0 & 0 & 0.5 & 2b \\
		%	\hline
		O1 &  0.094(1) & 0.026(1) &  0.263(2) & 4e \\
		%	\hline
		O2 & 0.186(3) & 0.296(4) & -0.057(1) & 4e \\
		%	\hline
		O3 &  0.198(3) & 0.290(4) & 0.532(2) & 4e \\
		\hline
	\end{tabular}
	\caption{Crystal structure of Pr$_2$NiIrO$_6$ at 150 K from neutron refinement in the $P2_1/n$ space group with $a$=5.473(2)$\rm \AA$, $b$=5.661(3)$\rm \AA$, $c$=7.790(3)$\rm \AA$, $\beta$=90.03(2)$^{\circ}$.}
	\label{Pr_table}
\end{table}

%trim is left, bottom, right, top
\begin{figure}[tb]
	\centering         
	\includegraphics[trim=0cm 6.8cm 0.0cm 0cm,clip=true, width=1.0\columnwidth]{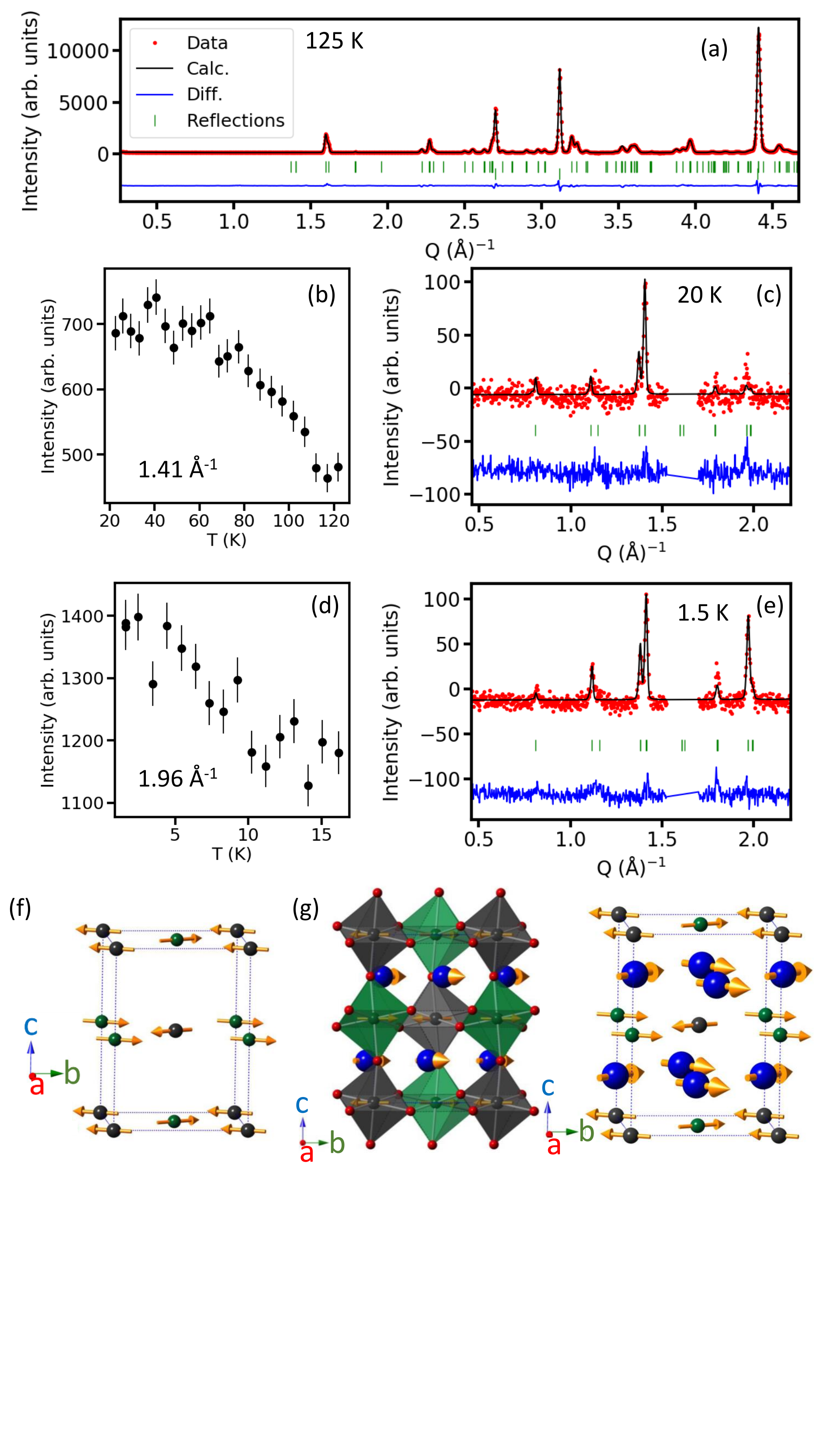}           
	\caption{\label{NPD_Pr} Magnetic structure of Pr$_2$NiIrO$_6$. (a) Refinement of the 125 K neutron diffraction pattern  to the crystal structure (upper reflections). Lower reflections correspond to Al sample holder scattering. (b) Intensity of the reflection at 1.41 $\rm \AA^{-1}$ as a function of temperature. (c) Magnetic structure model obtained by subtracting the 125 K neutron pattern from the 20 K data. Scattering around 1.6 $\rm \AA^{-1}$ due to the strong nuclear contribution. (d) Intensity of the reflection at 1.96 $\rm \AA^{-1}$ as a function of temperature. (e) Magnetic structure model obtained by subtracting the 125 K neutron pattern from the 1.5 K data. Scattering around 1.6 $\rm \AA^{-1}$ due to the strong nuclear contribution. (f) Magnetic-atom only representation of the magnetic structure at 20 K of Pr$_2$NiIrO$_6$ showing, Ni (green) and Ir (black) ions. The unit cell is outlined with the blue dashed line. (g) Polyhedral representation of the magnetic (1.5 K) and nuclear structure of Pr$_2$NiIrO$_6$ and  magnetic-atom only representation of the 1.5 K magnetic structure of Pr$_2$NiIrO$_6$ showing Pr (blue), Ni (green) and Ir (black) magnetic ions.}
\end{figure}

The composition Pr$_2$NiIrO$_6$ was measured at 1.5, 20, 75, and 125 K temperatures to follow anomalies observed in previous SQUID measurements \cite{doi:10.1021/acs.cgd.6b00121}. These indicated ferromagnetic-like ordering at 105 K with a further transition at 5 K. The high temperature 125 K neutron diffraction measurement shown in Fig.~\ref{NPD_Pr}(a) was used to confirm sample purity and the $P2_1/n$ structural model in the paramagnetic regime, see Table \ref{Pr_table}. Upon cooling below 110 K, additional Bragg reflections appeared. This is shown in Fig.~\ref{NPD_Pr}(b) by following the intensity at 1.41 $\rm \AA^{-1}$. Figure \ref{NPD_Pr}(c) shows all the observed magnetic reflections by subtracting the 125 K data from the 20 K data. There was no difference between the 20 K and 75 K measurements apart from increased intensity of the new magnetic reflections at the lower temperature measurement. Both temperatures have a  k  = (0, 0, 0) propagation vector. The scattering is similar to that observed for the Nd$_2$NiIrO$_6$ 20 K measurement, however here in the Pr$_2$NiIrO$_6$ case the signal to noise is improved and additional weaker reflections were observed. Following an identical analysis described above the magnetic space group of $P2_1/c (\#14.75)$ used for Nd$_2$NiIrO$_6$ was found to best fit the Pr$_2$NiIrO$_6$ data at 20 K, shown in Fig.~\ref{NPD_Pr}(c). The magnetic spins are primarily along the $b$-axis, however to model all the magnetic reflections a component along the $a$-axis needed to be added. This gives the ferrimagnetic structure shown in Fig.~\ref{NPD_Pr}(f) with the Ni ions ordered ferromagnetically and the Ir ions ordered ferromagnetically. Magnetic moments of 1.61(4)$\rm \mu_B/Ni^{2+}$ with components  ($\rm m_a$,$\rm m_b$,$\rm m_c$)= (0.6,1.5,0) and 0.34(8)$\rm \mu_B/Ir^{4+}$ with components  ($\rm m_a$,$\rm m_b$,$\rm m_c$)= (0.1,0.3,0) are found. Again the small moment size for Ir is presented as a best fit model and we cannot rule out a zero ordered moment. 

Cooling further from 20 K to 1.5 K additional magnetic scattering is observed as new intensity at certain reflections, while other positions such as at 1.38 $\rm \AA^{-1}$ and 1.41 $\rm \AA^{-1}$, remain unchanged. The intensity change at  1.96 $\rm \AA^{-1}$ is shown in Fig.~\ref{NPD_Pr}(d). Figure \ref{NPD_Pr}(e) shows all the observed magnetic reflections at 1.5 K by subtracting the 125 K data from the 1.5 K data. The propagation vector is also unchanged from the high temperature phase, k  = (0, 0, 0). This behavior is consistent with the ordering of the Pr ion while the Ni/Ir ions magnetic order remains unchanged. A clear contrast is observed with the Nd$_2$NiIrO$_6$ composition that showed a change in the Ni/Ir ordering at the low temperature magnetic phase transition. To model the 1.5 K data for Pr$_2$NiIrO$_6$ we keep the same magnetic space group of $P2_1/c (\#14.75)$ and include a moment on the Pr ion. The data refined to having the Pr ion in the $ab$-plane in a ferromagnetic arrangement similar to the Ni/Ir ions. The refinement to the 1.5 K data with the 125 K data subtracted is shown in  Fig.~\ref{NPD_Pr}(e) and the corresponding spin model in  Fig.~\ref{NPD_Pr}(g). The best fit model corresponds to magnetic moments of 1.63(4)$\rm \mu_B/Ni^{2+}$ with components  ($\rm m_a$,$\rm m_b$,$\rm m_c$)= (0.6,1.5,0) and 0.39(7)$\rm \mu_B/Ir^{4+}$ with components  ($\rm m_a$,$\rm m_b$,$\rm m_c$)= (0.1,0.3,0) and 1.58(3)$\rm \mu_B/Pr^{3+}$ with components  ($\rm m_a$,$\rm m_b$,$\rm m_c$)= (1.0,1.2,0). 

The onset of Pr ordering therefore contributes to the overall ferrimagnetic ordering within Pr$_2$NiIrO$_6$, with the Pr 1.58(3)$\rm \mu_B$ ordering ferromagnetically along the b axis in a zig-zag fashion due to a spin angle of 48.1(1)$^{\circ}$ off the b axis. The best fit model indicates the Pr and Ni ordering with the spins in the same direction along the $b$-axis and the Ir in the opposite direction. Further measurements on single crystals and with elemental specific analysis available with resonant x-ray scattering will be of interest to test this model and contrast it against a fully ferromagnetic ordering of all three magnetic ions. 

\subsection{RIXS measurements of SOC-Induced t$_{2g}$ manifold splitting}

%trim is left, bottom, right, top
\begin{figure}[tb]
	\centering         
	\includegraphics[trim=0cm 16.5cm 0.0cm 0cm,clip=true, width=0.8\columnwidth]{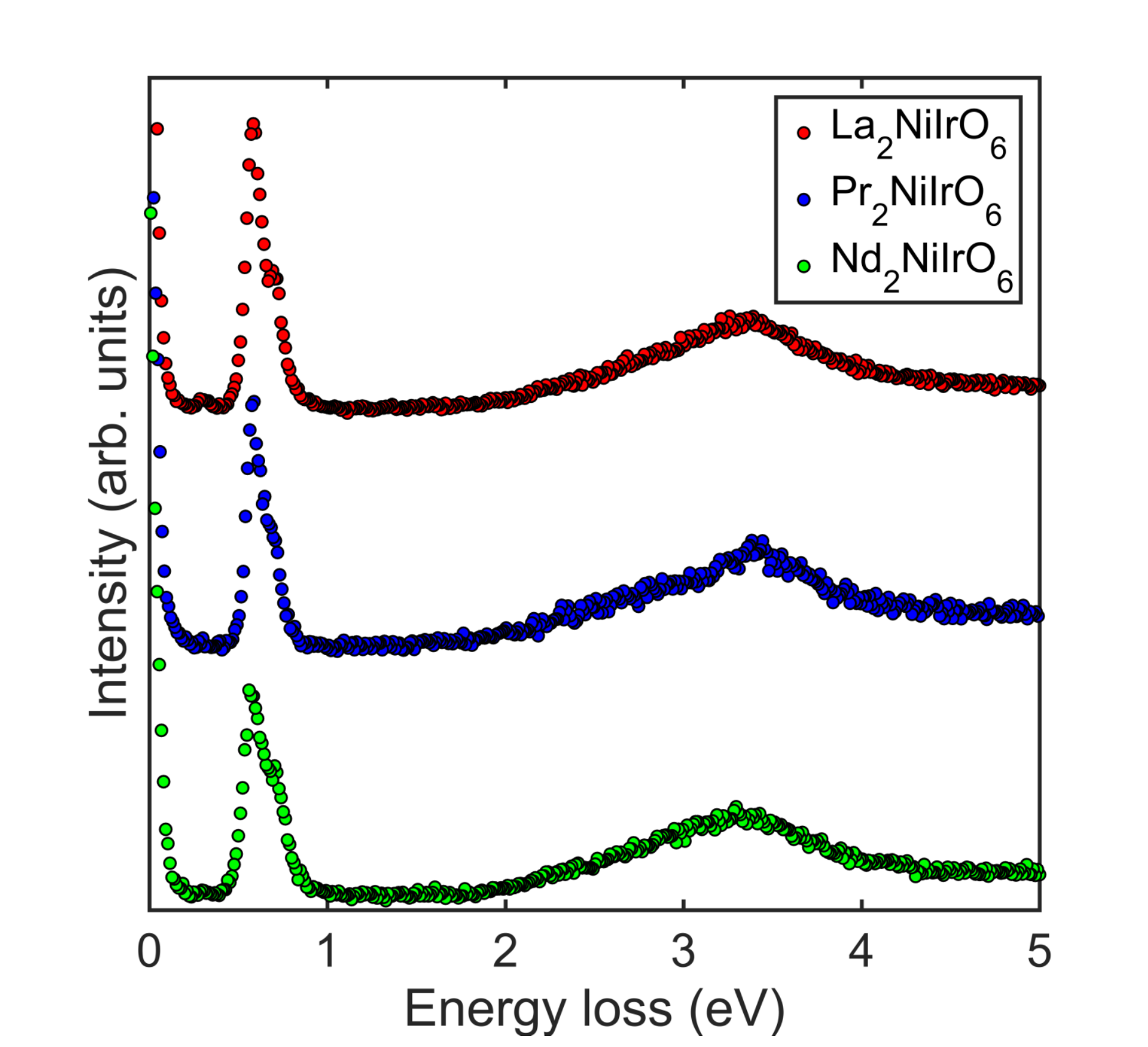}           
	\caption{\label{RIXS} Resonant inelastic x-ray scattering measurements of powder $Ln$$_2$NiIrO$_6$ ($Ln$ = La, Pr, Nd) at 5 K on the MERIX spectrometer. The incident energy was 11.215 keV corresponding to the Ir L$_3$-edge. The data have been offset by a constant factor for clarity.}
\end{figure}

To gain insight into the electronic ground state of the Ir$^{4+}$ (5$d^5$) ion in $Ln$$_2$NiIrO$_6$ RIXS measurements were performed. The energy was tuned to 11.215 keV corresponding to the Ir L$_3$-edge which allows for an isolation of the Ir scattering. The 5 K data is shown in Fig.~\ref{RIXS}. No change was observed in measurements collected at higher temperatures. Each compounds spectra consisted of two main features around 0.6 eV and 3.5 eV. The spectra are consistent with similar octahedrally coordinated Ir$^{4+}$ ion and provides all the signatures of a SOC split J$\rm _{eff}$ = $\frac{1}{2}$ state \cite{PhysRevB.99.134417,PhysRevB.100.085139}. The broad, higher energy (3.5 eV) peak corresponds to $d$-$d$ excitations from transitions between the t$_{2g}$ and e$_g$ orbitals, which are split because of the crystal field. The sharper, lower energy scattering consists of two separate peaks, within the 35 meV resolution of the instrument. This scattering can be assigned to $d$-$d$ excitations from intraband  t$_{2g}$ transitions due to the splitting of the t$_{2g}$ manifold into a J$\rm _{eff}$ = $\frac{1}{2}$ and J$\rm _{eff}$ = $\frac{3}{2}$ state, characteristic of many complex iridates \cite{PhysRevLett.102.017205}. By fitting these reflections to simple Gaussian peaks we extract peak energies as: La$_2$NiIrO$_6$= 0.60(2) eV and 0.71(2) eV; Pr$_2$NiIrO$_6$= 0.58(1) eV and 0.65(1) eV; Nd$_2$NiIrO$_6$= 0.59(2) and 0.71(3) eV. The presence of two resolvable peaks is consistent with a small departure from an ideal J$\rm _{eff}$ = $\frac{1}{2}$ state due to the distortions inherent in the crystal structure that is observed in all reported Ir materials in the literature \cite{PhysRevB.99.134417, PhysRevB.100.085139}. The singular unpaired electron present in the J$\rm _{eff}$ = $\frac{1}{2}$ level for 5$d^5$ Ir$^{4+}$ is commonly observed as possessing a significantly reduced magnetic moment, such as that observed in $Ln$$_2$NiIrO$_6$ reported above of $\sim$$0.3$$\rm \mu_B$, further supporting a J$\rm _{eff}$ = $\frac{1}{2}$ state.

\subsection{Heat Capacity, electrical resistivity and DFT investigations of Pr$_2$NiIrO$_6$}

\subsubsection{Heat Capacity of Pr$_2$NiIrO$_6$}

%trim is left, bottom, right, top
\begin{figure}[tb]
	\centering         
	\includegraphics[trim=0cm 2cm 0.0cm 0.0cm,clip=true, width=0.85\columnwidth]{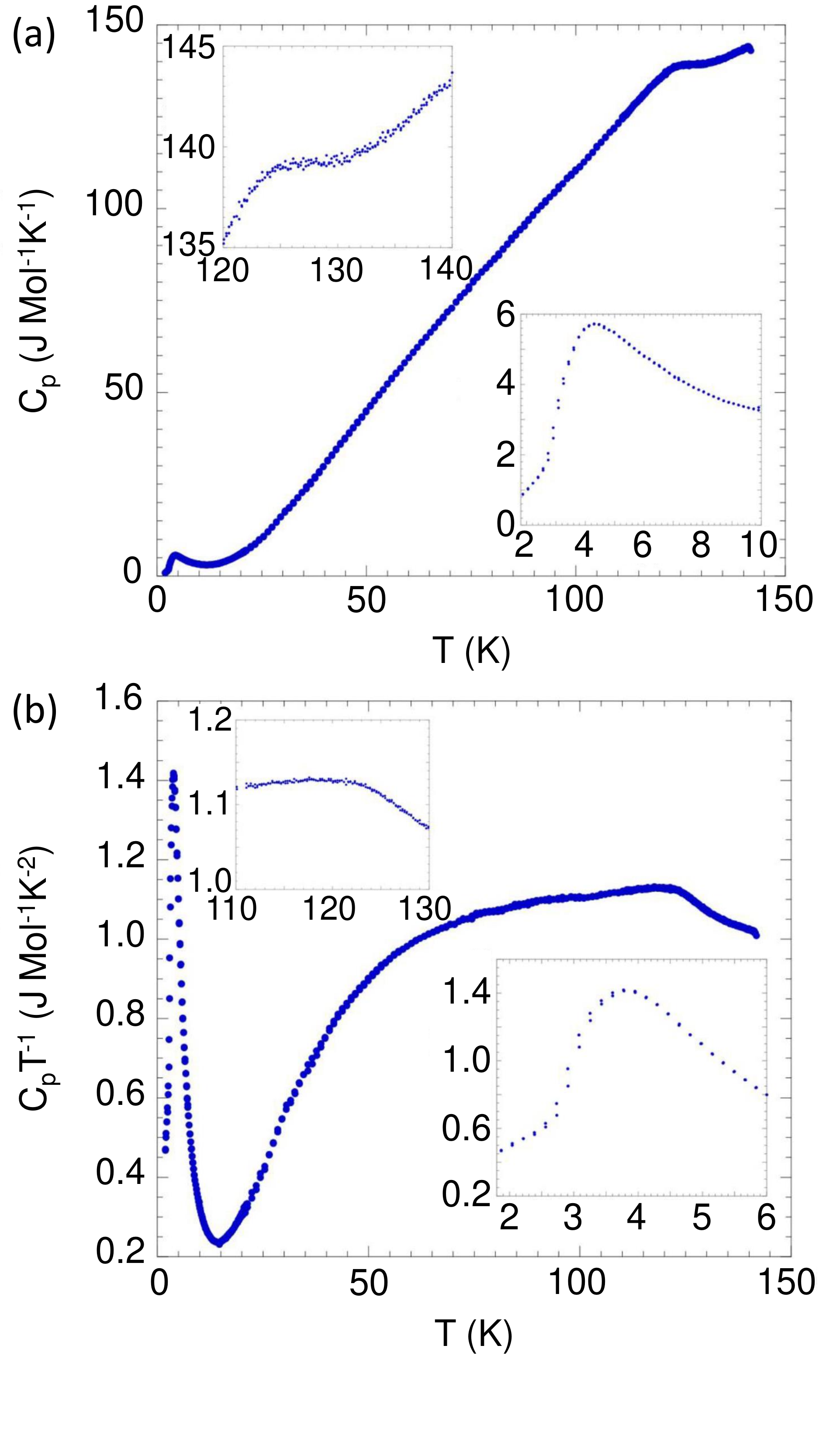}           
	\caption{\label{HeatCapacity} (a) Bulk heat capacity (C$\rm _p$) for Pr$_2$NiIrO$_6$ in zero field. The onset of Ni/Ir ordering is show	more clearly in the upper left inset, and the onset of Pr ordering is shown in the bottom right inset. (b) Bulk heat capacity divided by temperature (T) plotted against temperature for Pr$_2$NiIrO$_6$.}
\end{figure}

Heat capacity measurements were undertaken on a pressed and sintered pellet of Pr$_2$NiIrO$_6$, shown in Fig.~\ref{HeatCapacity}(a), to further investigate the long-range ordering and probe for independent Pr and Ni/Ir magnetic sublattices. Two clear transitions are observed at 123 K and 3.7 K, confirming the nature of long-range ordering temperatures. The broadness of the high temperature transition may be due to thermal fluctuation, a product of measuring at high temperature, or may be due to poor sintering of this sample. In addition we cannot rule out this as indicating low dimensional correlations for one or more of the ions. This transition was further resolved by plotting heat capacity (C$\rm _p$) divided by temperature (C$\rm _p$/T) against temperature, shown in Fig.~\ref{HeatCapacity}(b). Although broad features are still present, the 123 K transition is clear. As this transition is consistent with both neutron and susceptibility measurements indicating the onset of ferromagnetic-like order, its magnetic origin corresponds to the onset of Ni/Ir magnetic ordering.  Interestingly, the transition at 3.7 K was found to be  sharp and lambda-like, and is consistent with the small transition observed in zero-field cooled measurements shown in Fig.~\ref{HeatCapacity}(b) and the 1.5 K powder neutron diffraction data, suggesting the onset of Pr magnetic ordering. A small change in slope of heat capacity data can be observed below 2.7 K, but the nature of this transition is unclear. Based on the neutron and susceptibility data measurements, it does not correspond to any long range nuclear or magnetic order, suggesting possible crystal field effects between the three present magnetic ions in this structure. 

\subsubsection{Electrical Resistivity of Pr$_2$NiIrO$_6$}

%trim is left, bottom, right, top
\begin{figure}[tb]
	\includegraphics[trim=0cm 3cm 0.0cm 0cm,clip=true, width=0.85\columnwidth]{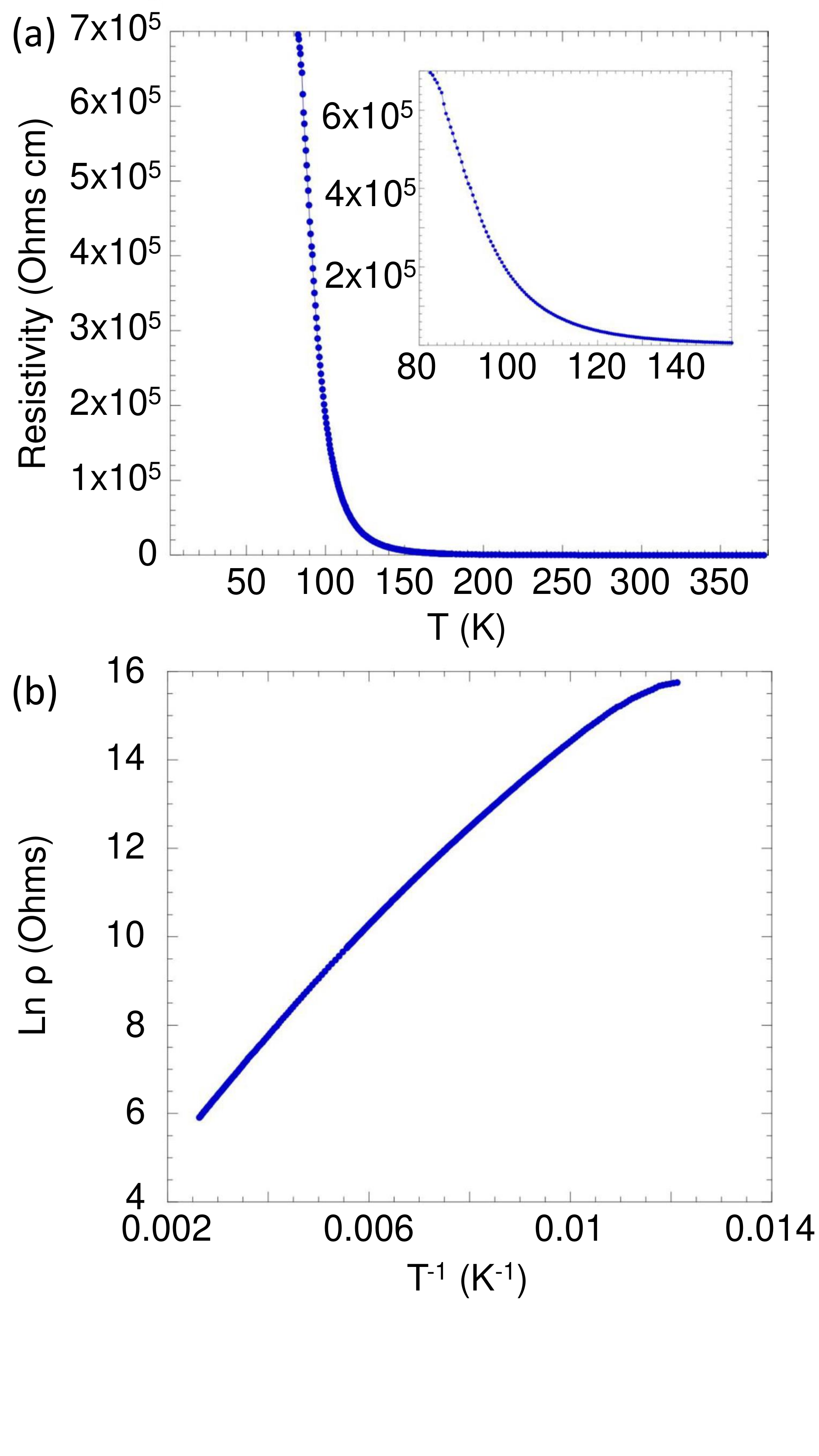}           
	\caption{\label{Resistivity} (a) Temperature dependence of the electrical resistivity for Pr$_2$NiIrO$_6$. The inset depicts that resistivity was not measured below 82.5 K due to instrumental limits of Ohms-cm capability. (b) Inverse temperature dependence of the natural logarithm of resistance for Pr$_2$NiIrO$_6$. Linearity for the case of T-n, such that n = 1, indicates thermally activated conduction.}
\end{figure}

Temperature dependent electrical resistivity measurements for Pr$_2$NiIrO$_6$ are shown in Fig.~\ref{Resistivity}(a). The sharp decrease in resistance as a function of increasing temperature indicates that the material is not metallic. Considering low measured resistance of 37 Ohms-cm at 380 K, especially for oxide materials  \cite{C7DT02254A}, semiconducting behavior is possible. This was further investigated by assessing the conduction mechanism via plotting the natural logarithm of resistance against T-n, such that the value of n indicates the dimensionality and type of transport mechanism. For values of n = 1, linearity indicates a simple thermally activated conduction pathway, whereas values for n greater than 1 indicate a Mott variable range hopping mechanism of variable dimensionality. Fig.~\ref{Resistivity}(b) depicts near perfect linearity is exhibited for n = 1, indicating the conduction pathway is thermally mediated.

\subsubsection{First Principles Calculations for Pr$_2$NiIrO$_6$}

In an attempt to better understand the complex magnetic behavior for Pr$_2$NiIrO$_6$, we have performed first principles calculations of the magnetic order and energetics. Given both the complex monoclinic physical structure as well as the non-collinear canted magnetic structure, with effectively 3 different magnetic ions (the dominant Ni, less dominant Pr, and induced moment Ir), we make certain simplifications in order to render the problem computationally and analytically tractable. First, we consider only collinear states.  While the actual observed ground state in Pr$_2$NiIrO$_6$ is not collinear, a detailed examination of Fig.~\ref{NPD_Pr}(f) (Ni/Ir ordering) and Fig.~\ref{NPD_Pr}(g) (all ions order) shows that deviations of the respective magnetic ions from collinearity are less than 30$^{\circ}$ off the $b$ axis for both Ni and Ir, but is a bit more significant for Pr. 

Given the monoclinic symmetry, there is a large manifold of potential exchange interactions, with several potential nearly-“nearest-neighbor” interactions, with slightly variable distances between Ni and Ni, Ni and Pr, Pr and Pr, and these atoms with Ir. To simplify matters we consider only Ni-Ni, Ni-Pr and Pr-Pr effective exchange interactions and consider the several nearly degenerate distances in each of these categories into one interaction for each category. We note in passing that it is not surprising that this compound exhibits a complex non-collinear magnetic structure in view of the complex physical structure and the three effectively magnetic ions, along with the disparate spin-orbit energy scales of Ni ($\sim$50 meV), Pr ($\sim$0.5 eV), and Ir ($\sim$1 eV). Note that the Ir atom is explicitly included in the DFT calculations themselves but for simplicity is not included in the extraction of
exchange constants as this would substantially complicate the analysis.

%trim is left, bottom, right, top
\begin{figure}[tb]
	\centering         
		\includegraphics[trim=0cm 16cm 0.0cm 0cm,clip=true, width=0.85\columnwidth]{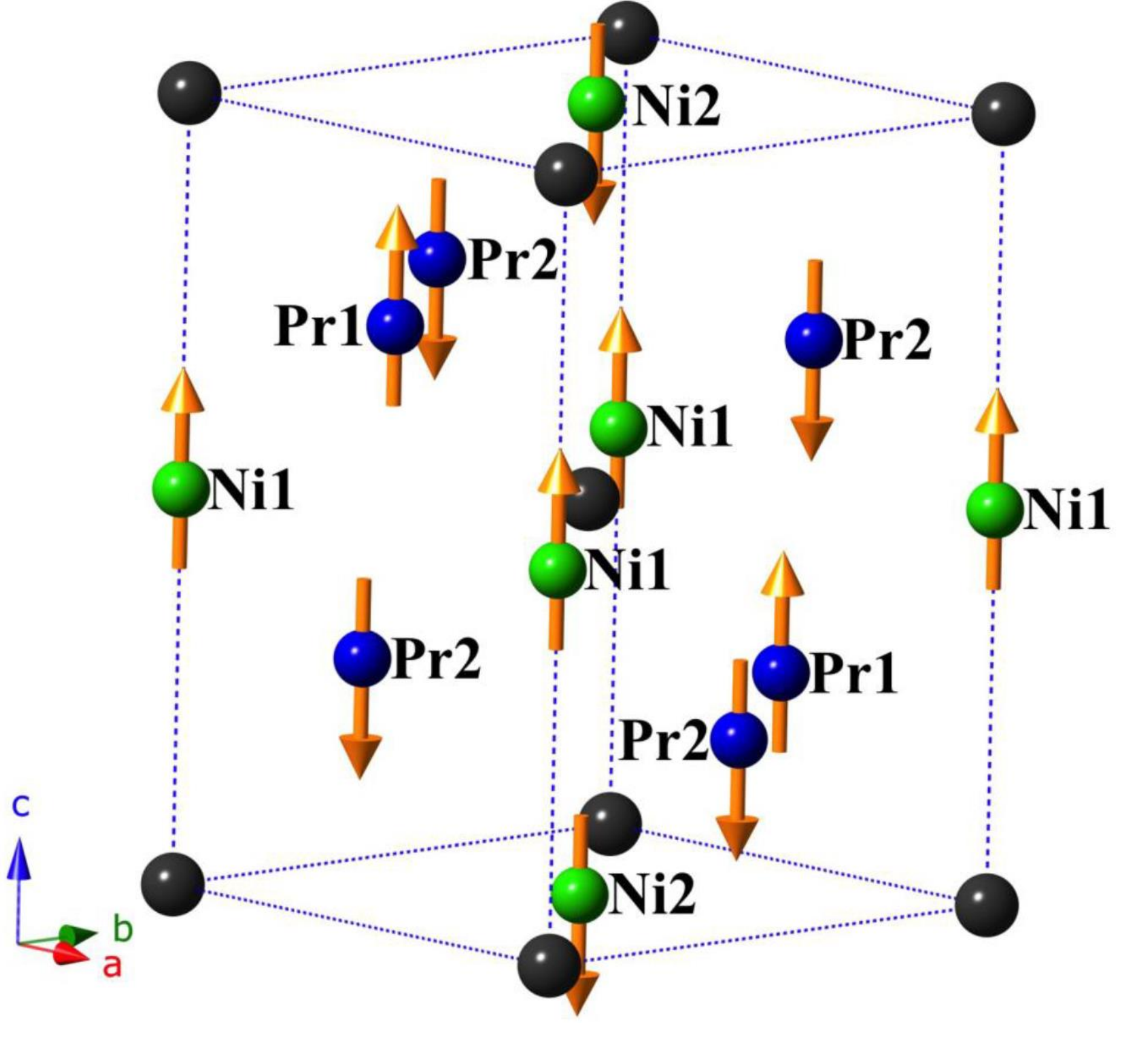}           
	\caption{\label{DFT_Fig} A depiction of the primary magnetic moment-bearing atoms described within the first principles calculations. Praseodymium atoms are depicted as blue spheres, Nickel atoms as green spheres, and Iridium atoms as dark grey spheres (no label), as indicated. For clarity, all spheres are shown as the same size, regardless of atomic size. The “Ni$\_$AF$\_$Pr$\_$AF” state is shown above. For the “FM” state, all Ni and Pr atoms are ferromagnetically coupled. For “Ni$\_$Pr$\_$FI”, the Ni and Pr atoms are antiferromagnetically coupled. For “Ni$\_$AF$\_$Pr$\_$FM”, the Ni1 and Ni2 atoms are antiferromagnetically coupled while all Pr atoms are ferromagnetically coupled to Ni1. For “Ni$\_$FM$\_$Pr$\_$AF”, Ni1 and Ni2 are ferromagnetically coupled while Pr1 is ferromagnetically coupled to Ni1 and Ni2 while Pr2 is antiferromagnetically coupled to Ni1 and Ni2. For “Ni$\_$AF$\_$Pr$\_$AF”, Ni1 and Ni2 are antiferromagnetically coupled and Pr1 and Pr2 are also antiferromagnetically coupled. Note that the {\it apparent} fourfold Pr2 falls on the $a$-face zone boundary (unlike Pr1, which is within the cell), so that there are only two Pr2 per unit cell. Similarly, the four Ni1 atoms fall on the zone edge, so that there is only one Ni1 per unit cell, and the two Ni2 atoms fall on the $c$-face zone boundary. The vertical moment orientation is for clarity of presentation; moment orientation was not studied in these calculations.}
\end{figure}

For the purposes of determining the ground state and associated excited state energetics, five distinct magnetic arrangements were considered. We show in Fig.~\ref{DFT_Fig} $–-$ a configuration with Ni-Ni and Pr-Pr near-neighbor pairs antialigned (Ni$\_$AF$\_$Pr$\_$AF). We considered four additional arrangements, including a ferromagnetic state (FM) and three more complex arrangements. For these purposes spin-orbit coupling was omitted, though for a detailed examination of the FM ground state below we include it. Here a crystallographic unit cell contains 2 formula units. The arrangements considered, in addition to the ferromagnetic case, were as follows: a state with Ni and Pr antiparallel to each other –a ferrimagnetic state (Ni$\_$Pr$\_$FI;l total moment 6 $\rm \mu_B/u.c$), a state with the two unit cell Ni antiparallel, with the Pr themselves aligned (Ni$\_$AF$\_$Pr$\_$FM, total moment 7.82 $\rm \mu_B/u.c$); a state with the 2 Ni ferromagnetically coupled, but 2 of the 4 Pr anti-aligned to the other 2; (Ni$\_$FM$\_$Pr$\_$AF, total moment 2 $\rm \mu_B/u.c$) and a state with the 2 Ni antiferromagnetically coupled, and 2 of the 4 Pr antiferromagnetically coupled to the other two (Ni$\_$AF$\_$Pr$\_$AF, no net moment). Note that of these last 4 states, only the last is truly a zero-moment antiferromagnetic state. In general, individual spin moment magnitudes within these several magnetic states are generally fairly rigid, with little variation ($<$ 3$\%$) between states; typical values, in the absence of spin-orbit coupling and thereby orbital moments, are 1.96 $\rm \mu_B/Pr$ and 1.29 $\rm \mu_B/Ni$. Details are given in Table \ref{DFTtable}. We will see below that despite the smaller moment and fewer atoms per cell, it is the Ni atoms that are ultimately dictating most of the magnetic character, due to the generally much larger spatial extent of the Ni 3$d$ wavefunctions, relative to the Pr 4$f$ wavefunctions, which are much more localized in the Pr core.

\begin{table}[tb]
	\begin{tabular}{|c|c|c|c|}
		\hline
	           & Energy          & Energy          &  Total \\
State  & relative to FM  & relative to FM  &  Spin Moment  \\
& GGA             & GGA+ U          &  ($\rm \mu_B$/u.c.,   \\
&                 & (per u.c.)      &  GGA+U)   \\
		\hline
		FM	&  0.0 & 0.0  & 10.0   \\
		\hline
		Ni$\_$Pr$\_$FI	& 49.56 &  41.96  &  6.0   \\
		\hline
		Ni$\_$AF$\_$Pr$\_$FM & 51.41 &  114.86  &    7.82       \\
		\hline
		Ni$\_$FM$\_$Pr$\_$AF & 103.48 &   37.68  &     2.0        \\
		\hline
		Ni$\_$AF$\_$Pr$\_$AF & 86.38 &   86.40  &     0.0        \\
		\hline   
	\end{tabular}
	\caption{Detailed magnetic properties of several magnetic configurations of Pr$_2$NiIrO$_6$ studied within density functional theory. The configurations’ relative orientation of the Ni and Pr spin magnetic moments are described in the text.}
	\label{DFTtable}
\end{table}

We see that from Table \ref{DFTtable}, the state with the 2 Ni atoms antiferromagnetically coupled, but the Pr ferromagnetically coupled, is the highest energy state in this manifold, nearly 115 meV/u.c. above the ferromagnetic ground state, while the reverse (Ni$\_$FM$\_$Pr$\_$AF) is only 37.7 meV/u.c. above. This accords with our intuitive expectation that Ni-Ni exchange interactions should be stronger than Pr-Pr exchange interactions, but what is surprising is that this is the case even though the Ni-Ni nearest neighbor distances are of the order of 5.7 $\rm \AA$ whereas those for Pr are only of order 4.1 $\rm \AA$. This is reflective both of the general localization of the Pr 4f electrons within the core, away from the Fermi level, and also of recent findings in Cr$_{1/3}$NbS$_2$  \cite{SiricaCommPhys} where exchange interactions mediated through electronegative elements can be much more long range than would commonly be expected. Quantitatively, mapping the above energetics to a Heisenberg model (appropriate in view of the rigidity of the moments) finds Ni-Ni, Ni-Pr and Pr-Pr exchange interactions of -4.32, -3.18 and -0.58 meV, (all ferromagnetic) respectively, confirming the expectation for the dominance of the Ni magnetic interaction here. In particular, the Pr-Pr exchange interaction is relatively weak and confirms our general expectation that the Pr 4$f$ electrons are localized in the core and do not interact strongly with other Pr atoms, reducing the Pr ordering temperature. 

Also evident from Table \ref{DFTtable} are substantially altered energetics in the ``straight GGA" calculations, in which no Hubbard U is applied to Pr. The exchange energetics change considerably; for example, the Ni$\_$FM$\_$Pr$\_$AF state in the straight GGA is now 103.48 meV/u.c. above the ferromagnetic state, whereas in the GGA+U it is
just 37.68 meV/u.c. above the unit cell, and this state now falls considerably higher (62 meV/u.c.) in energy above the Ni$\_$AF$\_$Pr$\_$FM state where it is some 77 meV/u.c. lower 
in the GGA+U. Most strikingly, extracting from these energetics the Ni-Ni, Ni-Pr and Pr-Pr exchange interactions, one now finds the Pr-Pr exchange interaction predominant in
magnitude at -8.65 meV, with the Ni-Pr exchange at -4.05 meV and the Ni-Ni much smaller at just -0.06 meV. Thus the straight GGA would here inaccurately claim the Pr-Pr 
magnetic interaction to be the predominant one, a logical consequence of the GGA's placing the Pr 4f orbitals at or near the Fermi level, where they interact strongly with other atoms, instead of being properly localized in the core of the Pr atom, as a wealth of experience with rare earth ions dictates. Thus the application of a substantial U value (here chosen
as 5 eV) is critical to a proper description of the magnetism in this compound.

As mentioned previously, we now give a more complete description of the FM ground state with spin-orbit coupling included for all atoms with the GGA+U approach. This changes both the total spin moment significantly (it increases to 10.68 $\rm \mu_B$/u.c.) and adds significant orbital moment contributions. In particular, Pr exhibits a large negative orbital moment of -0.795 $\rm \mu_B$, while Ni acquires a significant orbital moment of 0.087 $\rm \mu_B$, and Ir also has a significant negative orbital moment of -0.197 $\rm \mu_B$. This yields total moments for these three atoms of 1.17 $\rm \mu_B$, 1.45 $\mu_B$, and -0.35 $\rm \mu_B$ with a unit cell total of 7.28 $\rm \mu_B$, or 3.64 $\rm \mu_B$/f.u. The above values are comparable to the experimental values of 1.58(3) $\rm \mu_B$ for Pr, 1.63(4) $\rm \mu_B$ for Ni, and 0.39(7) $\rm \mu_B$ for Ir. It is of interest that the largest orbital moment magnitudes are for the heavy atoms (Pr and Ir), corresponding to the generally stronger spin-orbit coupling in these atoms, as well as the localized nature of the Pr 4$f$ states. The addition of spin-orbit also changes the Iridium spin-moment from -0.30 $\rm \mu_B$ to a value half this, suggesting the particular relevance of spin-orbit coupling here, as evidenced experimentally in the RIXS measurements discussed above.

It is of interest both that the Iridium atom couples antiferromagnetically to the Pr and Ni atoms and that its spin moment is significantly reduced by the application of spin-orbit coupling. Sitting at the center of a distorted Oxygen octahedron, its antiferromagnetic coupling is notable in view of the ferromagnetic coupling of Ni itself, sitting near the
center of a similar Oxygen octahedron. It is also remarkable that its orbital moment of -0.197 $\mu_B$ is significantly larger in magnitude than its spin moment of -0.15 $\mu_B$, and this may be understood in terms of its spin-orbit energy scale of $\sim$ 1 eV effectively outstripping its exchange interaction energy scales, which may be expected to be much
smaller than the predominant Ni-Ni exchange interaction. Note that the Ir orbital and spin moments are parallel, in accordance with Hund's rules.  It is likely that the Oxygen atoms intervening between the Ni and Ir atoms cause a predominant antiferromagnetic superexchange interaction between these atoms.  The Ir atom thus likely plays an important
role in the overall Ni-Ni magnetic exchange interaction, despite carrying a comparatively small moment itself.  The total calculated Ir moment of 0.35 $\mu_B$ is consistent  both with the neutron-found value of $\sim$ 0.3 $\mu_B$ and with the J= 1/2 state inferred from the RIXS measurements.

To further relate these results to the RIXS data we present in Figure 8 the calculated density-of-states (DOS), within the modeled ferromagnetic state, with spin-orbit coupling in the GGA+U. For simplicity we present only the majority spin (or spin-up) DOS and focus on the RIXS-relevant region around the Fermi level. One immediately notes the presence of substantial, in fact predominant Ir character(blue dotted line) in the region within half an eV of E$_{F}$. Furthermore, we note two Ir-generated DOS peaks falling approximately -0.25 below and 0.15 eV above E$_{F}$. It is quite likely that 0.4 eV transitions between these regions correspond to the Ir intraband t$_{2g}$ transitions observed at $\sim$ 0.6 eV in the RIXS,  with the difference in energies (0.4 vs. 0.6 eV) reasonably ascribed to our simplified treatment of the magnetism in this system.

\begin{figure}[tb]
	\includegraphics[width=0.85\columnwidth]{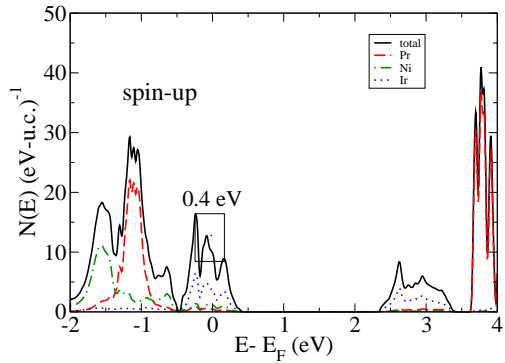}           
	\caption{The calculated spin-up density-of-states of Pr$_2$NiIrO$_6$ in the modeled ferromagnetic groundstate, within GGA+U with spin-orbit coupling applied. Note the RIXS-relevant transition indicated around the Fermi level, indicative of the intraband Ir transition found in RIXS}
\end{figure}

It is also possible to relate the above GGA+U energetics to the observed ordering points of the Ni and Pr atoms, which significantly differ in temperature. In an approximation where the ordering point of a magnetic atom, in a local moment approximation, is estimated at 1/3 the energy difference, per magnetic atom, \cite{pbe, doi:10.1063/1.4961933,PhysRevApplied.9.034002}, between configurations with that atom ferromagnetically and antiferromagnetically coupled to the remainder of the system, we use the Ni$\_$FM$\_$Pr$\_$AF state for the Pr atom (relative to the FM ground state) and, correspondingly, the Ni$\_$AF$\_$Pr$\_$FM for the Ni atom, accounting for the different multiplicity of these atoms, and obtain estimated ordering points for the Pr and Ni atoms as 36 and 222 K. While these are somewhat higher than the actual values (due to our wholesale neglect of fluctuations, among other factors), their relative magnitudes are in accordance with the experimental facts. In particular, as seen in experiment, the Nickel atoms are driving the magnetism, despite their roughly equivalent local moment and substantial nearest-neighbor distances. This mainly reflects the spatially extended nature of the 3$d$ states associated with the Ni atom magnetism and the generally core-localized nature of the Pr 4$f$ electrons associated with the Pr magnetism.

\section{Conclusions}

A series of double perovskite iridates of the formula $Ln$$_2$NiIrO$_6$  were investigated using thermodynamic and transport properties, neutron powder diffraction,  RIXS, and DFT calculations to elucidate the role superexchange plays in hybrid 3$d$-5$d$-4$f$ compositions with variable A-site cations. The composition La$_2$NiIrO$_6$ was determined to be a non-collinear antiferromagnet in the $ac$-plane with the ordering of Ni/Ir occurring simultaneously. For Nd$_2$NiIrO$_6$ two distinct magnetic structures were determined. The first high temperature magnetic phase consists of ferromagnetically ordered Ni/Ir sublattices creating a ferrimagnetic structure primarily along the $b$-axis. Cooling Nd$_2$NiIrO$_6$ led to a magnetic structure change where the Nd ion orders and the Ni/Ir ordering also changes into a AFM Ni/Ir sublattices. Two independent magnetic sublattices (Pr and Ni/Ir) were found in the composition Pr$_2$NiIrO$_6$, corresponding to $ab$-plane ferrimagnetic order between Ni and Ir, and a zig-zag ferromagnetic order of Pr along the $b$-axis, resulting in an overall ferrimagnetic order. The presence of two independent magnetic sublattices was corroborated by heat capacity measurements, demonstrating transitions at 123 K (Ni/Ir ordering) and 3.7 K (Pr ordering). Resistivity measurements indicated semiconducting behavior and thermally mediated conduction for Pr$_2$NiIrO$_6$. DFT results confirm the independent sublattice ordering and demonstrate the primacy of the Ni atom in determining the magnetic character, despite the Ni-Ni nearest-neighbor distances of some 5.7 $\rm \AA$. All compositions were measured with RIXS, confirming that spin-orbit coupling splits the t$_{2g}$ manifold of octahedral Ir$^{4+}$ into a J$\rm _{eff}$ = $\frac{1}{2}$ and J$\rm _{eff}$ = $\frac{3}{2}$ state. Collectively the results demonstrate the dramatic changes in magnetic ordering that can be induced within structurally similar 3$d$-5$d$-4$f$ compounds as the $Ln$ ion is varied and as different temperature regimes are accessed. As shown in the DFT calculations and experimental data the presences of distinct magnetic ions with a spectrum of SOC strength and orbital overlaps leads to the inducing of magnetic interactions that otherwise would not occur and goes beyond predictions that apply to simpler systems less magnetic ions, such as the Kanamori-Goodenough rules. This motivates further investigations into hybrid materials with multiple magnetic ions.

\begin{acknowledgments}
	
 This research used resources at the High Flux Isotope Reactor, a DOE Office of Science User Facility operated by the Oak Ridge National Laboratory. This work was partly supported by the U.S. Department of Energy (DOE), Office of Science, Office of Workforce Development for Teachers and Scientists, Office of Science Graduate Student Research (SCGSR) program. The SCGSR program is administered by the Oak Ridge Institute for Science and Education for the DOE under contract number DE-SC0014664. Research at Oak Ridge National Laboratory (ORNL) was supported by the DOE, Office of Science, Basic Energy Sciences (BES), Materials Science and Engineering Division. Sample synthesis and structural characterization performed at the University of South Carolina were supported by the National Science Foundation under Awards DMR-1301757 and DMR-1806279. This work used resources of the Advanced Photon Source, an Office of Science User Facility operated for the U.S. DOE Office of Science by Argonne National Laboratory. This manuscript has been authored by UT-Battelle, LLC under Contract No. DE-AC05-00OR22725 with the U.S. Department of Energy. The United States Government retains and the publisher, by accepting the article for publication, acknowledges that the United States Government retains a non-exclusive, paidup, irrevocable, world-wide license to publish or reproduce the published form of this manuscript, or allow others to do so, for United States Government purposes. The Department of Energy will provide public access to these results of federally sponsored research in accordance with the DOE Public Access Plan(http://energy.gov/downloads/doepublic-access-plan).

\end{acknowledgments}

%\bibliography{Dperov_refs}

%merlin.mbs apsrev4-1.bst 2010-07-25 4.21a (PWD, AO, DPC) hacked
%Control: key (0)
%Control: author (72) initials jnrlst
%Control: editor formatted (1) identically to author
%Control: production of article title (-1) disabled
%Control: page (0) single
%Control: year (1) truncated
%Control: production of eprint (0) enabled
%

\end{document}